\documentclass[%
reprint,
superscriptaddress,
 amsmath,amssymb,
 aps,
 prl,
]{revtex4-2}

\usepackage{graphicx}
\usepackage{dcolumn}
\usepackage{bm}
\usepackage{hyperref}
\usepackage[mathlines]{lineno}

\usepackage{xcolor}
\begin{document}

\preprint{APS/123-QED}

\title{Magic-angle Bilayer Phononic Graphene}

\author{Yuanchen Deng}
 \thanks{Y.D. and M.O. contributed equally to this work.}
    \affiliation{Graduate Program in Acoustics,  Penn State University, University Park, Pennsylvania, 16802, USA}
\author{Mourad Oudich}
 \thanks{Y.D. and M.O. contributed equally to this work.}
    \affiliation{Graduate Program in Acoustics,  Penn State University, University Park, Pennsylvania, 16802, USA}
    \affiliation{Université de Lorraine, CNRS, Institut Jean Lamour, F-54000 Nancy, France}
\author{Nikhil JRK Gerard}
    \affiliation{Department of Mechanical and Aerospace Engineering, North Carolina State University, Raleigh, North Carolina 27695, USA}
\author{Jun Ji}
    \affiliation{Graduate Program in Acoustics,  Penn State University, University Park, Pennsylvania, 16802, USA}    
\author{Minghui Lu}
    \affiliation{National Laboratory of Solid State Microstructures and Department of Materials Science and Engineering, Nanjing University, Nanjing, Jiangsu 210093, China}
    
\author{Yun Jing}
    \email{yqj5201@psu.edu}
    \affiliation{Graduate Program in Acoustics,  Penn State University, University Park, Pennsylvania, 16802, USA}
\date{\today}

\begin{abstract}

Thanks to the recent discovery on the magic-angle bilayer graphene, twistronics is quickly becoming a burgeoning field in condensed matter physics. This letter expands the realm of twistronics to acoustics by introducing twisted bilayer phononic graphene, which remarkably also harbors the magic angle, evidenced by the associated ultra-flat bands. Beyond mimicking quantum mechanical behaviors of twisted bilayer graphene, we show that their acoustic counterpart offers a considerably more straightforward and robust way to alter the interlayer hopping strength, enabling us to unlock magic angles ($>3^{\circ}$) inaccessible in classical twisted bilayer graphene. This study, not only establishes the acoustical analog of twisted (magic-angle) bilayer graphene, providing a test bed more easily accessible to probe the interaction and misalignment between stacked 2D materials, but also points out the direction to a new phononic crystal design paradigm that could benefit applications such as enhanced acoustic emission and sensing.  
\end{abstract}

\maketitle


Van der Waals (vdW) heterostructures vastly expand the family of 2D materials and have been a central topic in materials physics~\cite{Geim2013VanHeterostructures, Ponomarenko2013CloningSuperlattices, Gong2014VerticalMonolayers, Gorbachev2014DetectingSuperlattices, Ajayan2016Two-dimensionalMaterials}. Twisted bilayer graphene (TBG), which entails two graphene sheets placed on top of each other with a small angle misalignment, has served as an emerging theoretical and experimental platform to study vdW heterostructures owing to their intriguing electronic and optical properties   ~\cite{Luican2011Single-layerLayers, LopesDosSantos2007GrapheneStructure, LopesDosSantos2012ContinuumBilayer}. This field of research concerning how the twist between layers of 2D materials can alter and tailor their electronic behavior was coined “twistronics” ~\cite{Carr2017Twistronics:Angle}. Recent experiments on TBG have made ground-breaking discoveries on correlated (Mott) insulating ~\cite{Cao2018CorrelatedSuperlattices} and unconventional superconducting states~\cite{Cao2018UnconventionalSuperlattices}. At the heart of these findings lies the existence of flat electronic bands near the Fermi energy, when the twist angle is close to the so called “magic” angles ~\cite{Bistritzer2011MoireGraphene, Tarnopolsky2019OriginGraphene, Bistritzer2011MoireGrapheneb, LopesDosSantos2012ContinuumBilayer, Cao2018CorrelatedSuperlattices, Cao2018UnconventionalSuperlattices, Yankowitz2019TuningGraphene,Nam2017LatticeGraphene}. These flat bands exhibit insulating states at half-filling, a characteristic that can only be explained by electron-electron interactions. 
The initial experimental work on magic-angle bilayer graphene has spurred a proliferation of studies, which have further provided crucial and complimentary findings pertaining to magic angles, such as tunable superconductivity~\cite{Yankowitz2019TuningGraphene}, Kohn-Luttinger superconductivity ~\cite{Gonzalez2019Kohn-LuttingerGraphene}, nontrivial topological phases of magic angles~\cite{Song2019AllTopological}, emergent geometric frustration~\cite{Pal2019EmergentGraphene}, and charge order and broken rotational symmetry in magic-angle bilayer graphene~\cite{Jiang2019ChargeGraphene}. 
 
Simultaneously over the past few years, artificial materials such as photonic and phononic crystals have become a fertile playground for mimicking quantum-mechanical features of condensed matter systems and have revealed new routes to controlling classical waves~\cite{He2016AcousticTransport,Deng2017ObservationDefects,Zhu2018SimultaneousSystem,Wang2019ExceptionalSystem,Wang2018BoundCoupling}. Following the path of building analogues to topological and Chern insulators ~\cite{Yang2015TopologicalAcoustics,He2016AcousticTransport, Ozawa2019TopologicalPhotonics, Ma2019TopologicalSystems,Xue2019AcousticLattice,Ni2019ObservationSymmetry}, valley Hall effects~\cite{Lu2016ValleyCrystals,Lu2017ObservationCrystals}, Weyl semimetals~\cite{Xiao2015SyntheticSystems,Li2018WeylCrystal,Xie2019ExperimentalCrystal}, and Landau levels ~\cite{Schine2019ElectromagneticLevels,Wen2019AcousticStates} in classical wave systems, some recent works have attempted to introduce vdW heterostructures and twistronics to acoustics~\cite{Dorrell2020VanMetamaterials, Lu2018ValleyCrystals} and optics~\cite{Wang2019ObservationGraphene,Wang2020LocalizationLattices,Hu2020MoireMetasurfaces}. Nevertheless, the direct analogue of TBG as well as that of magic angles has not been studied in phononic systems. Additionally, these existing designs present feasibility constraints for tuning the interlayer hopping strength  (i.e., coupling strength in acoustics), a key parameter directly linked to the magic angle. Driven by the potential of twist-enabled acoustic energy localization and new topological physics brought about by the magic angles, we study the acoustic version of magic angles for a twisted bilayer phononic graphene (TBPG). The proposed twisted phononic platform also offers an extraordinarily simple approach for radically changing the interlayer hopping strength, allowing us to engineer a wide range of magic angles not accessible in classical TBG~\cite{Yankowitz2019TuningGraphene,Carr2018PressureSuperlattices}. Specifically, this study demonstrates two magic angles greater than $3^{\circ}$, which is the upper bound of the experimentally accessible magic angle in TBG under uniaxial pressure~\cite{Carr2018PressureSuperlattices}. 

In order to realize the equivalent of a bilayer graphene where the upper layer eigenstates interact with the ones in the lower layer, we begin first by building the equivalent of the monolayer graphene for the case of acoustic waves~\cite{Torrent2012AcousticWaves,Yu2016SurfaceGraphene}. The monolayer phononic graphene is shown in FIG. 1(a). It consists of a rigid plate with a hexagonal lattice of air cavities of cylindrical shape (or air columns). The unit cell contains two cavities which form two identical air columns that are inter-connected via the air above them. The distance between the two air columns is $a_0=10$ mm,  rendering a lattice constant $|\vec a_1| = |\vec a_2| =a= \sqrt3 a_0$. The length of the air columns is 20 mm and the diameter is 7.2 mm. The air columns in this structure serve as acoustic ``atoms" which create a Dirac cone in the band structure (FIGs. 1(b) and (c), shown at 3838 Hz), reminiscent of what is observed in monolayer graphene.  The band structure is obtained by COMSOL Multiphysics 5.4. The shaded region covers the area above the sound line, which separates the spoof surface acoustic wave (SSAW) from the open space acoustic modes. Owing to the cavity resonance, the eigenstates of the Dirac point (FIG. 1(d)) are associated with a very low group velocity, which is manifested by the flat bands in the vicinity of the Dirac point (FIG. 1(b)). Such eigenstates show characteristics similar to those of surface acoustic waves in elastodynamics and therefore the corresponding wave is known as the spoof surface acoustic wave ~\cite{Kaina2015NegativeMetamaterials,Wu2019RoutingAnisotropy,Liu2018UnidirectionalCrystal}, which is evanescent in the direction normal to the rigid plate.

\begin{figure}
\includegraphics[width=8.5cm]{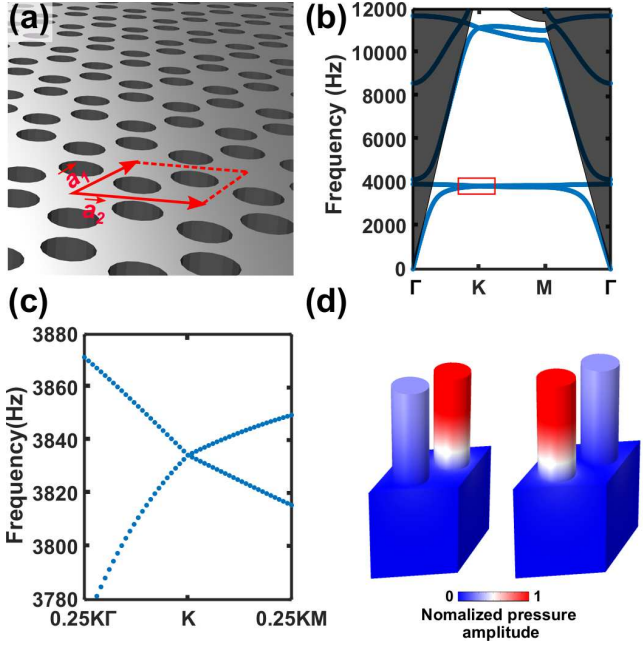}
\caption{\label{fig:1}(a) The monolayer phononic graphene made of air cavities on a rigid plate. $\vec a_1$ and $\vec a_2$ are the lattice vectors. (b) The band structure of the monolayer phononic graphene. The shaded area represents the frequency range above the sound line and the red box indicates the Dirac point at the K point. (c) Enlarged view in the vicinity of the Dirac point. (d) The degenerate eigenstates at the Dirac point at $3838$ Hz.}
\end{figure}

 The bilayer phononic graphene is then assembled by stacking up two of such monolayer structures while leaving an air-gap of thickness $h$ in between (FIG. 2(a) and FIG. 3(b)). In this manner, the coupling between the SSAWs hosted by the two monolayer phononic graphene is analogous to the interlayer hopping effect in bilayer graphene. When one of the phononic graphene is twisted, the resulting system shows a Moir\'e pattern manifested by a periodic arrangement of AA, AB and BA stacking regions (FIG. 3(a)) \cite{Cao2018CorrelatedSuperlattices}. We first investigate the AA and AB stacking where the unit cells are presented in FIG. 2(a). The band structures of these unit cells are studied in order to estimate the interlayer hopping strength, denoted $w$ (termed interlayer hopping energy in bilayer graphene~\cite{Bistritzer2011MoireGraphene}). For the case of AA stacking with $h=15$ mm, the band structure shows a pair of Dirac cones at the $K$ point (FIG. 2(b)). Whereas in the case of AB (or Bernal) stacking with the same $h$, a parabolic-like dispersion of the Dirac bands appears (FIG. 2(c)). In both cases, the band structures are similar to those observed in bilayer graphene~\cite{Rozhkov2016ElectronicSystems}.

\begin{figure}
\includegraphics[width=8.5cm]{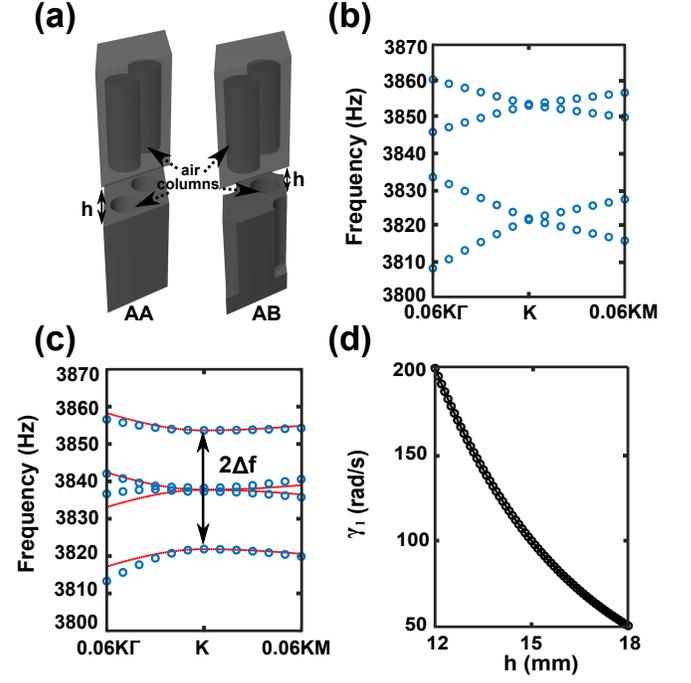}
\caption{\label{fig:2} (a) Unit cells of the TBPG for the AA (left) and AB (right) stacking. (b) The band structure of the AA stacking near the K point. (c) The band structure of the AB stacking near the K point. circles: numerical results; solid lines: TBM results.  (d) The interlayer vertical hopping $\gamma_1$ is shown as a function of $h$.}
\end{figure}

 We further adopt a tight-binding model (TBM) to gain insight on the band structure of the phononic bilayer graphene in the vicinity of the K point. In the case of AB stacking, the Hamiltonian can be written in a way similar to that of AB-stacked bilayer graphene~\cite{McCann2013TheGraphene}, which yields
\begin{equation}
\begin{pmatrix}
\omega_{0}& -\gamma_{0}f(\mathbf{k}) &\gamma_{4}f(\mathbf{k}) & -\gamma_{3}f^*(\mathbf{k})  \\ 
-\gamma_{0}f^*(\mathbf{k}) & \omega_{0} & \gamma_{1} & \gamma_{4}f(\mathbf{k}) \\ 
\gamma_{4}f^*(\mathbf{k}) &   \gamma_{1} & \omega_{0}& -\gamma_{0}f(\mathbf{k}) \\
-\gamma_{3}f(\mathbf{k}) & \gamma_{4}f^*(\mathbf{k}) & -\gamma_{0}f^*(\mathbf{k}) & \omega_{0} \\
 \end{pmatrix} ,
\end{equation}
where
\begin{equation}
    f(\mathbf{k})=e^{ik_y a/\sqrt{3}}+2e^{-ik_y a/2\sqrt{3}}\cos(k_x a/2).
\end{equation}

The on-site energy $\omega_{0}$ is chosen as the degeneracy frequency. Following the notion of the Slonczewski--Weiss--McClure
(SWM) model, $\gamma_0$ describes the in-layer hopping whereas $\gamma_1$,$\gamma_3$, and $\gamma_4$ are the interlayer hopping terms. Notably, $\gamma_1$ describes the interlayer vertical hopping and $\gamma_1=2\pi\Delta f$ (FIG. 2(c))   ~\cite{Kuzmenko2009DeterminationSpectroscopy,Rozhkov2016ElectronicSystems}. We consider only the first-order interlayer hopping by treating $\gamma_3$ and $\gamma_4$ as zero~\cite{Dorrell2020VanMetamaterials}. 
For more details on the physical meaning of $\gamma_0$, $\gamma_1$,$\gamma_3$, and $\gamma_4$, the reader is referred to Fig. 2 of reference~\cite{McCann2013TheGraphene}. The effective Dirac velocity of the monolayer phononic graphene $v$ is given by $v=a\gamma_0\sqrt{3}/2$ and is estimated to be around 4.1 m/s through fitting (See supplemental material~\cite{SuppMaterial}). The TBM-derived band structure is shown in FIG. 2(c).

In contrast to previous bilayer phononic designs that are based on having an interlayer made of membranes ~\cite{Dorrell2020VanMetamaterials} or perforated plates~\cite{Lu2018ValleyCrystals}, the design proposed here offers the possibility of tuning the interlayer hopping strength $w$ by simply adjusting the air-gap thickness $h$. Figure 2(d) shows how $\gamma_1$, which is proportional to $w$~\cite{Bistritzer2011MoireGraphene}, varies with the thickness $h$. There is a significant drop of $\gamma_1$ as the thickness $h$ increases. This is fully anticipated since the coupling of the SSAWs weakens as the two phononic graphene move further apart. 

\begin{figure}
\includegraphics[width=8.5cm]{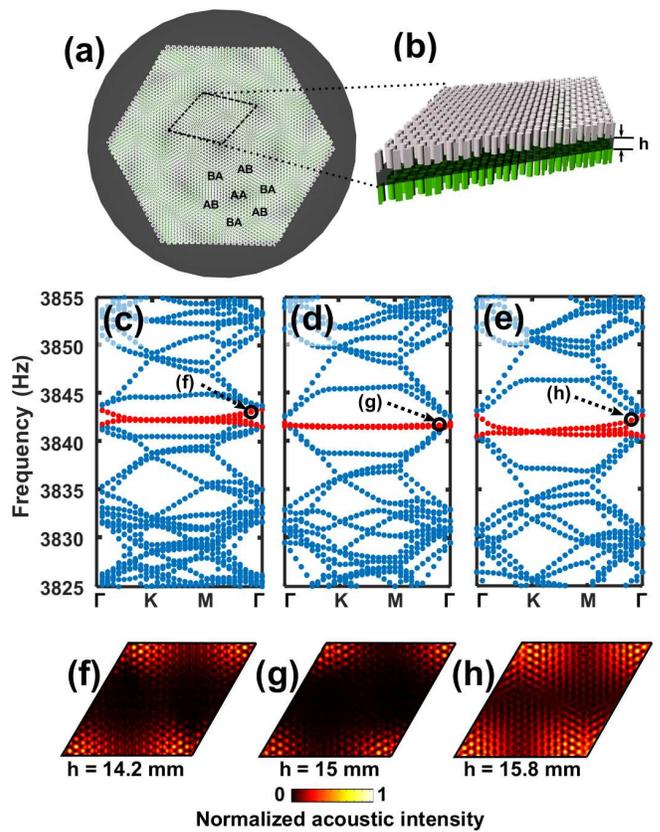}
\caption{\label{fig:3}(a) The top view of the TBPG with a twist angle of $3.481^{\circ}$. The Moir\'e pattern is indicated by the alternating dark and bright regions of the TBPG which correspond to AA, AB and BA stackings. (b) The side view of the supercell, where air columns are shown and $h$ represents the thickness of the inter-connected air-gap. (c-e) Band structures of three TBPG samples with $h$ of $14.2$ mm, $15$ mm and $15.8$ mm, respectively. The red lines highlight the evolution of the flat bands.(f-g) The eigenmode acoustic intensity distributions are shown for the circled points for each TBPG.}
\end{figure}

The magic angle in bilayer graphene originates from interlayer hybridization, which induces isolated flat bands ~\cite{Tarnopolsky2019OriginGraphene, Bistritzer2011MoireGraphene}. This was first demonstrated in the electronic dispersion of TBG and an analogy can hence be drawn to the TBPG system. Specifically, we will use the fact that the magic angle is accompanied by flattened bands near the Dirac point frequency, whose bandwidth (BW) is minimum at the $\Gamma$ point~\cite{Tarnopolsky2019OriginGraphene}. In this study, instead of varying the twist angle which is common practice in TBG, we first fix the twist angle at 3.481$^{\circ}$. This is specifically a commensurate angle that gives rise to strictly periodic superlattices \cite{SuppMaterial}. Consequently, the band structure of this TBPG can be computed by exact wave-based methods in COMSOL and the BW can be extracted. Figures 3(c-e) show the band structures of three TBPG possessing different interlayer hopping strength. Their air-gap thickness $h$ are 14.2 mm, 15 mm, and 15.8 mm, respectively. A flat band at around 3842 Hz can be clearly seen in the 15 mm TBPG, whereas the bands open up at the $\Gamma$ point in the other two cases. A close-up view reveals that the flat band in the 15 mm TBPG in fact encompasses four flat bands (supplemental material~\cite{SuppMaterial}), which is similar to that in the magic-angle bilayer graphene~\cite{Cao2018CorrelatedSuperlattices}. These four bands are evolved from the two Dirac cones of the top and bottom layer graphene. Figures 3(f-h) show the corresponding eigenmode acoustic intensity at the circled points in the momentum space for the three samples, respectively. The 15 mm TBPG shows the strongest localization of energy, a hallmark of flat bands. Interestingly, the energy is localized around the AA stacking regions at the four corners. This is similar to magic-angle bilayer graphene where the local density of states peaks at the AA stacking region~\cite{Cao2018CorrelatedSuperlattices}. Figure 4 further plots the BW as a function of $h$ (red circles), which shows that the BW indeed reaches the minimum at 15 mm. 

Theoretically, the BW in TBG can be predicted by the following equation~\cite{Tarnopolsky2019OriginGraphene}, 
\begin{equation}
\label{BW}
    BW =  \frac{2w}{\alpha}\times (1-2\alpha+\frac{\alpha ^{2}}{3}+\frac{2\alpha ^{3}}{9}+\frac{4\alpha ^{4}}{54}+...), 
\end{equation}
where $\alpha$ is related to the twist angle $\theta$ by
\begin{equation}
    \alpha = w/vk_{\theta}.
\end{equation}
$k_{\theta}$ is related to the separation between the two Dirac cones that are to be hybridized in the Brillouin zone. The hybridization therefore relies on $\theta$ and $k_{\theta}$ which can be expressed as  $k_{\theta} = 2k_{D}\sin(\theta/2)$, where $k_D$ is the magnitude of the Brillouin-zone corner wave vector for the monolayer graphene. It is assumed in TBPG that $w = A  \gamma_1$, where $A$ is a fitting parameter and $\gamma_1$ can be obtained from FIG. 2(d). After proper conversion from energy to frequency following the quantum-acoustic analogue (Eq. \ref{BW} divided by $2\pi$), the theoretically predicted BW of TBPG is shown in FIG. 4 (red line), with $A=0.35$. Note that this value of $A$ is reasonable as it falls in the range ($0.33$ - $0.40$) reported by previous studies in TBG~\cite{Bistritzer2011MoireGraphene, LopesDosSantos2012ContinuumBilayer}.     
We have also studied a second case with a larger commensurate angle of 5.086$^{\circ}$. In theory, a twist angle smaller than 3.481$^{\circ}$ can be also investigated, such as around 1.1$^{\circ}$ in TBG~\cite{Cao2018CorrelatedSuperlattices}. This is not done in the present study since the corresponding TBPG would have a very large supercell that cannot be handled by the computational resources available to the authors. The BW plot of the TBPG at 5.086$^{\circ}$ is given in FIG. 4 (blue), where the same $A$ value is used. The band structures and eigenmodes can be found in the supplemental material~\cite{SuppMaterial}. Numerical results indicate that the flat bands emerge when the air-gap thickness $h$ is 13.6 mm. The slight deviation between the theory and numerical result can be possibly attributed to the fact that the flattened bands at the $\Gamma$ point are not symmetrical, which violates the assumption made in the theoretical model~\cite{Tarnopolsky2019OriginGraphene}. This asymmetry of bands, which is also evident in FIG. 2(c), seems to be intrinsic for SSAWs as it was not observed in conventional acoustic waves~\cite{Dorrell2020VanMetamaterials}. Finally, we have also obtained the band structure for a 3.150$^{\circ}$ 
TBPG with $h=15$ mm. This band structure is used to illustrate the evolution of the flat bands as the twist angle increases while the interlayer coupling maintains~\cite{SuppMaterial}.

\begin{figure}
\includegraphics[width= 7.5 cm]{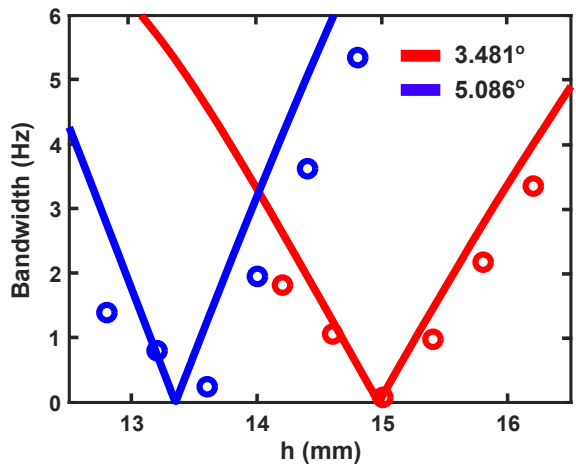}
\caption{\label{fig:4} Bandwidth of the flat bands at the $\Gamma$ point as a function of air-gap thickness $h$. The 3.481$^{\circ}$ TBPG results are shown in red color whereas the 5.086$^{\circ}$ TBPG results are shown in blue color. Solid lines represent theoretical results whereas circles represent numerical results.}
\end{figure}

In conclusion, we have developed a design paradigm for phononic crystals that exploits twist and interlayer coupling as new degrees of freedom.
The proposed TBPG offers a distinct macroscopic platform for twistronics, where prior efforts have been limited to quantum systems at the atomic scale. We demonstrate two previously inaccessible magic angles at 3.481$^{\circ}$ and 5.086$^{\circ}$, where flat bands emerge near the Dirac point frequency. These large magic angles entail the advantage of reduced overall size of the sample, which is vital for the miniaturization of magic-angle-inspired devices. Furthermore, the eigenmodes at these flat bands show strong localization of acoustic energy, which could be proven useful for enhancing acoustic emission and sensing. The proposed TBPG can be readily constructed for experimental investigation (see supplementary material for the effect of loss), paving the way for future research on vdW heterostructures and twistronics in the realm of acoustics. Finally, we envision that the results presented here can inspire new designs of twisted photonic and elastic-wave 2D materials, extending their impact throughout the bosonic system.

\begin{acknowledgements}
This work was supported in part by NSF through CMMI-1951221. 
\end{acknowledgements}


%


\end{document}